\documentclass[12pt,letterpaper]{article}

\usepackage{graphicx,array}
\usepackage{url}
\usepackage{color}
\usepackage{latexsym}
\usepackage{amsthm}
\usepackage{amsmath}
\usepackage{amssymb}
\usepackage{amsfonts}
\usepackage[numbers,sort&compress]{natbib}
\usepackage{bm}
\usepackage{slashed}
\usepackage{mathrsfs}
\usepackage{enumerate}
\usepackage{tikz}
\usepackage{siunitx}
\usepackage{mdframed}
\usepackage{setspace}  
\usepackage{esvect}
\usepackage{esint}
\usepackage{multirow}
\usepackage{multicol}
\usepackage{tablefootnote}

\usepackage[utf8x]{inputenc}%
\usepackage{tcolorbox}%

\usepackage{hyperref} 
\hypersetup{
    colorlinks=true,       
    linkcolor=red,          
    citecolor=blue,        
    filecolor=magenta,      
    urlcolor=blue           
}
\usepackage[all]{hypcap} 

\usepackage{natbib}
\def\GeV{\mathrm{GeV}}     

\newcommand{\eg}{{\it e.g.}}
\newcommand{\ie}{{\it i.e.}}

\usepackage{floatrow}
\newfloatcommand{capbtabbox}{table}[][\FBwidth]

\usepackage{blindtext}

\title{ 
{\bf Searching for Magnetic Monopoles with the Earth's Magnetic Field}
\author{\large Yang Bai$^{\,\star}$, Sida Lu$\,^\diamond$, and Nicholas Orlofsky$^{\,\dagger}$}
\date{\small \it 
$^\star$Department of Physics, University of Wisconsin-Madison, Madison, WI 53706, USA\\
$^\diamond$Department of Physics, Tel Aviv University, Tel-Aviv 69978, Israel \\
$^\dagger$Department of Physics, Carleton University, Ottawa, ON K1S 5B6, Canada \\
}
}

\begin{document}

\maketitle

\setlength{\parskip}{0.2ex}

\begin{abstract}
Magnetic monopoles have long been predicted in theory and could exist as a stable object in our universe. As they move around in galaxies, magnetic monopoles could be captured by astrophysical objects like stars and planets. Here, we provide a novel method to search for magnetic monopoles by detecting the monopole moment of the Earth's magnetic field. Using over six years of public  geomagnetic field data obtained by the {\it Swarm} satellites, we apply Gauss's law to measure the total magnetic flux, which is proportional to the total magnetic charge inside the Earth. To account for the secular variation of satellite altitudes, we define an altitude-rescaled magnetic flux to reduce the dominant magnetic dipole contribution. The measured magnetic flux is consistent with the existing magnetic field model that does not contain a monopole moment term. We therefore set an upper limit on the magnetic field strength at Earth's surface from magnetic monopoles to be $|B_{\rm m}| < 0.13$\,nT at 95\% confidence level, which is less than $2\times 10^{-6}$ of Earth's magnetic field strength. This constrains the abundance of magnetically-charged objects, including magnetic black holes with large magnetic charges. 
\end{abstract}

\thispagestyle{empty}  
\newpage  
\setcounter{page}{1}  
%
%

\section{Introduction}\label{sec:Introduction}

As observed by Dirac close to a century ago~\cite{Dirac:1931kp}, the magnetic monopole is a fascinating physical object that could elegantly explain the quantization of electric charges in Nature. Ever since, physicists have been studying magnetic monopoles both from theoretical and experimental directions. On the theory side, the discovery of Polyakov-'t Hooft monopoles in 1974~\cite{Polyakov:1974ek,tHooft:1974kcl} has been applied to Grand Unified Theories (GUT)~\cite{Georgi:1974sy}, which predict the GUT monopole mass around $10^{17}$~GeV. For heavier masses above the Planck mass scale, magnetically charged black holes have long been proposed, which can have masses proportional to their magnetic charges~\cite{Lee:1991qs,Lee:1991vy,Maldacena:2020skw}. 
Various experimental methods have been adopted to search for magnetic monopoles, \eg, detecting the quantized jump in magnetic flux when monopoles pass a superconducting quantum interference device~\cite{Cabrera:1982gz}
and searching for the Cherenkov light generated when the accelerated monopoles pass the large IceCube detector~\cite{Aartsen:2015exf}. Magnetic monopoles could also be captured by stars and planets including our Earth, and their annihilations can produce detectable neutrinos and/or heat~\cite{Carrigan:1980an,Frieman:1985dv,Bai:2020spd} (see also \cite{Bai:2020spd,Mavromatos:2020gwk,Ghosh:2020tdu,Diamond:2021scl} for further constraints). In this article, we present a new way to search for magnetic monopoles by measuring the magnetic monopole moment of the Earth.

Ever since the classical 1839 paper by Gauss~\cite{Gauss1877} (see~\cite{hgss-5-11-2014} for English translation), measurements indicate the Earth's magnetic field is dominantly dipole-like with subleading contributions from higher moments. Although the general spherical harmonics formula developed by Gauss contains the monopole moment~\cite{Gauss1877}, it has been assumed to be zero because the traditional Maxwell equations lack magnetically charged objects. As magnetic monopoles have a solid theoretical motivation and a chance to be captured inside the Earth, the Earth's magnetic monopole moment could be non-zero. Performing a measurement of Earth's monopole moment is therefore a straightforward and worthwhile method to search for magnetic monopoles. 

To measure different magnetic moments including the monopole moment of the Earth, one could perform a global fit like the one used for the International Geomagnetic Reference Field~\cite{alken:hal-02993250}. In this article, we adopt a simpler approach and apply Gauss's law to Earth's magnetic field to obtain the total magnetic flux, proportional to the enclosed magnetic charge. 
Ideally, one needs to perform synchronous measurements of magnetic field vectors at all points on a sphere surrounding Earth to obtain a precise value for the magnetic flux. In practice, ground-based measurements only cover a small fraction of the total area, although the measurements can be made synchronously. For satellite-based measurements, the satellites visit different locations at different times but do provide good whole-sky coverage. Knowing that the secular variation of magnetic fields is in general much longer than the satellite orbiting time (order of hours for one orbit and order of days for whole-sky coverage), we will use the monopole moment. More specifically, we will analyze the publicly available data by the {\it Swarm} satellites~\cite{Swarm-2006,Swarm-data} spanning more than 6 years of observation. For equations, we use natural units with $\hbar = c =\varepsilon_0 = 1$, although numerical values are expressed in the International System of Units (SI).

\section{Measuring the monopole moment}

\subsection{Measuring magnetic charges via Gauss's law}
For a single object or a group of objects with a total magnetic charge of $Q$ at the center of the Earth, the Earth's magnetic field has a monopole moment of $\bm{B}_{\rm m}(\mathbf{r})  =  \frac{Q\,h}{4\,\pi\,r^2} \,\mathbf{\hat{r}}= \frac{Q}{2\,e\,r^2} \,\mathbf{\hat{r}}$, where $e = \sqrt{4\pi \alpha}$ with $\alpha \approx 1/137$ as the fine-structure constant and $h=2\pi/e \approx 68.5\,e\approx 21$ is the magnetic coupling. $Q=1$ is the minimal magnetic charge, corresponding to the Dirac quantization condition with $e\,h = 2\pi$~\cite{Dirac:1931kp}. Numerically, $B_{\rm m} \approx 0.082 \,\mbox{nT} \times \left( \frac{R_\oplus}{r} \right)^2 \, \left(\frac{Q}{10^{19}}\right)$,
where $R_\oplus\approx 6371.2$~km is the average radius of the Earth. For comparison, the measured Earth surface intensity has a magnitude of up to $\approx 65000~\text{nT}$.

To measure the magnetic charge, one could adopt Gauss's law $\oint\, \bm{B}(\mathbf{r}) \cdot d\bm{A} = Q\, h$. This requires a full-sky measurement of the magnetic vector field. For convenience, one could choose the manifold to be a sphere of radius $R$ centered on Earth. Then, $d\bm{A} = R^2\,\mathbf{\hat{n}}\,d\Omega$ with $\mathbf{\hat{n}}$ as a unit surface vector pointing outward and $d\Omega = \sin{\theta}d\theta d\phi$ in spherical coordinates. For magnetic monopole objects, $\overline{B}_\text{m} \equiv \frac{1}{4\pi} \oint\bm{B}_{\rm m}(r, \theta, \phi) \cdot \mathbf{\hat{n}}\, d\Omega  = Q\,h \, \frac{1}{4\pi\,R^2}$.
Here, we have defined a solid-angle averaged magnetic field $\overline{B}$, which is simply the amplitude of the monopole magnetic field at radius $R$. Its sign matches the sign of the magnetic charge. 
All higher multiple moments beyond $\bm{B}_\text{m}$ do not contribute to $\overline{B}$. 

In practice, the measurement of magnetic field is not performed at a uniform radius---the {\it Swarm} satellite orbits have a variation of $\mathcal{O}(1\%)$ during one orbit and decay over time.
Thus, it is not possible to integrate the magnetic flux along a perfectly spherical closed manifold, and the surface's normal vector $\mathbf{\hat{n}}$ will not match the radial coordinate unit vector $\mathbf{\hat{r}}$. So, a numerical integration of $\int\bm{B}(r, \theta, \phi) \cdot \mathbf{\hat{r}} \, d\Omega$ will not be zero, even in the absence of a monopole term (for {\it Swarm}'s orbital parameters, $\overline{B} \simeq -70~\text{nT}$; see the Supplemental Material).
To suppress this measurement-induced dipole contribution, we use the following modified Gauss law to measure the magnetic field from the monopole charge
\begin{equation}
\label{eq:B-modified}
\overline{\mathscr{B}} =  \frac{1}{4\pi}  \int \, \left[\frac{r(\theta, \phi)}{R_{\rm ref}}\right]^3\, \bm{B}(r, \theta, \phi) \cdot \mathbf{\hat{r}}\, d\Omega ~.
\end{equation}
Here, $r(\theta, \phi)$ is the radius of the magnetic measurement at different angular directions and $R_{\rm ref}$ is a fixed reference radius.
For the dipole component, this is formally equivalent to integrating on a perfectly spherical surface at $r=R_\text{ref}$, so $\mathbf{\hat{n}}=\mathbf{\hat{r}}$ and the dipole component contributes zero to the above quantity. Note that the Earth's higher-moment magnetic fields have non-zero contributions to the quantity $\overline{\mathscr{B}}$  
because the the higher moments scale with higher powers of $r$.
For instance, the quadrupole moment has a magnitude of $\mathcal{O}(10\%)$ of the dipole moment, and contributes around 0.5~nT for $\overline{\mathscr{B}}$ using {\it Swarm}'s orbit.
Therefore, the $r^3$ scaling in Eq.~\eqref{eq:B-modified} is practically useful to improve the sensitivity of searching for the monopole moment because it reduces contributions to $\overline{\mathscr{B}}$ from the dipole and higher moments while preserving the monopole signal. 

\subsection{{\it Swarm} satellites}

The European Space Agency {\it Swarm} mission launched in late 2013. It consists of three satellites in low-Earth near-polar orbits each measuring the local magnetic field vector once per second. Two satellites {\it Swarm A} and {\it Swarm C} orbit in a similar plane with only about 1.5 degrees east-west separation and initial altitude 460 km, while {\it Swarm B} is separated from the other two and has initial altitude 530 km. The altitudes vary with time due to atmospheric drag (causing an overall decrease with time) and orbital eccentricity (changing the satellite radius by less than 1\% within each orbit). Because each orbit takes around 90 minutes and the Earth completes a rotation every 24 hours, each satellite can be thought of as sweeping out measurements at all latitudes in about 32 different longitudes per day. The orbital inclinations of the satellites are $87.4^\circ$ for {\it Swarm A} and {\it C} and $88^\circ$ for {\it Swarm B}, so that no measurements are taken within $2^\circ$ of the geographic poles.

We use the VirES architecture to access the data via the Python package \texttt{viresclient} \cite{ashley_smith_2021_4476534}. Magnetic field measurements are used from the {\it Swarm} L1b 1\,Hz data product. When selecting data, we take the following considerations. Due to a sensor failure in {\it Swarm C}, we only use the data from the other two satellites. Bad data is removed using the Flags\_F filter, and only non-zero vector data for which the magnetic activity level $Kp \leqslant 3$ are included to reduce the impact of the Sun (following the procedure in Refs.~\cite{Sabaka_swarm,Finlay_2020}). Unlike other analyses, we do not impose a cut on the Sun's elevation angle above the horizon, as this would prevent us from obtaining full $4\pi$ sky coverage (particularly near the poles).

\subsection{Data analysis and model comparison}
\label{sec:stat}

To calculate the flux, the measurements are binned into $a \times a$ degree angular patches and $d$-day-long time bins. If the satellites do not cover all $a \times a$ degree patches within a given $d$ day time bin (for example, due to the selection cuts), that bin is not used.  Because one satellite covers about 32 different longitudes per day, it takes at least $d \geqslant 360/(32a)$ days for one satellite to cover every patch, although generally it takes longer due to selection cuts and the precise orbital path taken. Because it takes about 45 minutes for a satellite to transit from the north to the south pole, it spends about $a/2$ minutes in each angular patch. With 1\,Hz measurement cadence, it takes about $30 a$ measurements per angular patch in a single orbit. Due to the orbital inclinations, the patches closest to the poles are imposed to have size $3^\circ$ (latitude) $\times\,a^\circ$ (longitude) when $a<3$.

We index the individual measurements, angular patches, and time bins by $i$, $j$, and $k$ respectively. To estimate the magnetic flux for a given time bin,
we first rescale the measurements of the magnetic field in the radial direction $B_r$  to a common radius $R_\text{ref} = 6850~\text{km}$, the average radius of both satellites for the mission duration thus far, as $B_{r,i}(\theta_i,\phi_i) = B_{r,i}(r_i, \theta_i,\phi_i) (r_i/R_\text{ref})^3$. Then, we average every rescaled measurement of the magnetic field in the radial direction within each angular patch $\left<B_r\right>_{j} = N_j^{-1} \sum_{i \mid (\theta_i,\phi_i) \, \in \, \text{patch }j} B_{r,i}(\theta_i,\phi_i)$, with $N_j$ the number of measurements in patch $j$.  These patch averages are then summed together, weighted by the area of each patch, to obtain 
$\overline{\mathscr{B}} = \sum_j \left<B_r\right>_{j}w_j$, with $w_j = (4 \pi)^{-1} \int_{ \text{patch }j} \sin \theta d\theta d\phi$. To obtain an estimate of the statistical error, the squared standard error of the mean is calculated for each angular patch as $\sigma_j^2 = (N_j-1)^{-1}N^{-1}_j \sum_{i \mid (\theta_i,\phi_i) \, \in \, \text{patch }j} [B_{r,i}(\theta_i,\phi_i) - \left<B_r\right>_j]^2$.  We ignore the errors in each measurement, which are $\lesssim 0.5~\text{nT}$, as trivial compared to the errors from having different measurements at different positions and times within each patch. These errors are weighted-summed in quadrature to obtain the squared error of the total flux $\sigma^2 = \sum_j \sigma^2_j w^2_j$. For each time bin $k$, this procedure yields an estimate for $\overline{\mathscr{B}}_k$ and its statistical uncertainty $\sigma_k$. The error for each time bin scales as $\sigma_k \propto a$. 

The approach outlined above contains several systematic errors, so $\overline{\mathscr{B}}$ cannot be identified with the magnetic flux $\overline{B}$.  These include the imperfect spherical coverage which depends on the orbital path taken, the finite angular patch size used to numerically integrate (\ref{eq:B-modified}), the finite time bin size when compared against the non-static nature of the fields, and the other magnetic field components besides the dominant dipole component which are not cancelled using the $r^3$ rescaling in (\ref{eq:B-modified}). 
Many of these systematic errors can be reduced by subtracting the data measurement $\overline{\mathscr{B}}_{k}^\text{dat}$ from the value $\overline{\mathscr{B}}_{k}^\text{mod}$ predicted by a model of Earth's field (that does not contain a monopole moment term) evaluated at the same coordinates $(r,\theta,\phi,t)$ of each satellite measurement. 
 
Models for Earth's magnetic field contain contributions from the dominant dipole moment term, higher moments coming from both the core and the lithosphere, and external sources including the ionosphere and magnetosphere field. 
The internal components are expressed as a spherical harmonic expansion with time-dependent coefficients.
We use the CHAOS-7 model~\cite{Finlay_2020}, which is fitted through 20 July 2020.

To search for a monopole, we compare how well the data agrees with the model, the latter of which contains no monopole moment term.  First, we calculate the difference in each time bin $\overline{\mathscr{B}}_k^\text{dif} = \overline{\mathscr{B}}_k^\text{dat} - \overline{\mathscr{B}}_k^\text{mod}$ and add their uncertainties in quadrature $(\sigma_k^\text{dif})^2 = (\sigma_k^\text{dat})^2 + (\sigma_k^\text{mod})^2$. 
Then, we perform a weighted average of $\overline{\mathscr{B}}_k^\text{dif}$ using
$\langle\overline{\mathscr{B}}^{\rm dif}\rangle = (\sigma^{\rm dif})^2 \sum_k \overline{\mathscr{B}}^{\rm dif}_k / (\sigma_k^{\rm dif})^2$. The squared error of $\langle\overline{\mathscr{B}}^{\rm dif}\rangle$ is $(\sigma^{\rm dif})^2 = [\sum_k 1/(\sigma_k^{\rm dif})^2]^{-1}$.
Where the $\chi^2$ difference between the time-binned measurements $\overline{\mathscr{B}}_k^\text{dif}$ and the mean $\langle\overline{\mathscr{B}}^\text{dif}\rangle$ indicates additional variability, we enlarge the total error $\sigma$ by the Birge ratio $r_{\rm B} = \sqrt{\chi^2/(N_k-1)}$~\cite{Birge-PhysRev.40.207,2007Metro..44..415L} or $\sigma = \mbox{\rm max}[1, r_{\rm B}]\, \sigma^\text{dif}$, with $N_k$ the number of time bins. When $r_B>1$, this enlarged error is essentially equivalent to the standard error of the mean of $\langle\overline{\mathscr{B}}^\text{dif}\rangle$, \ie, $(\sigma^{\rm sem})^2 = (N_k-1)^{-1}N^{-1}_k \sum_{k} (\overline{\mathscr{B}}_k^\text{dif} - \langle\overline{\mathscr{B}}^\text{dif}\rangle )^2$. 
Note, $\sigma^{\rm sem}$ is irreducible in the limit $a \rightarrow 0$ provided $a$ is small enough for each time bin's measured $\overline{\mathscr{B}}_k^\text{dif}$ to stabilize numerically (but not so small that there are empty angular patches without measurements). It can be interpreted as including additional variability effects not included in the purely statistical error, including different orbital paths or acceptance cuts in each time bin.~\footnote{If a significant signal were detected, the variability could also be due to the time evolution of the monopole moment as charged objects are captured by Earth, an effect we neglect in the absence of a signal.}

In the absence of a signal, we set a bound on the monopole charge by injecting an artificial signal $\bm{B}_\text{m}(\mathbf{r})$ into the data. 
A new $\langle\overline{\mathscr{B}}^\text{dif}\rangle$ is calculated by comparing the data+injected signal to the model, and the bound on the injected signal size is set where $\langle\overline{\mathscr{B}}^\text{dif}\rangle$ differs from zero at 95\% confidence level.

To provide a sense of how much each component of Earth's field contributes to our estimate of the flux, the various contributions are summarized in Table \ref{tab:model}, showing the prediction for $\langle\overline{\mathscr{B}}^\text{mod}\rangle$ (the time average of the model alone) from various components of the CHAOS-7 model~\cite{Finlay_2020} using $a=2$, $d=180$, and $Kp \leqslant 3$ (see also the shaded bands in Fig.~\ref{fig:result}). The core component is separated into the dipole and higher moments. The second column shows the calculation under idealized conditions where all measurements are taken at the same time and on an equal-radius sphere, with $(\theta, \phi)$ still matching the satellite measurements. These terms formally go to zero when $a \rightarrow 0$, so they can be interpreted as the systematic error from the discretization of the angular integral. 
The crust and external components contribute negligibly (see also Fig.~\ref{fig:crust_vs_ext} in the Supplemental Material).

\begin{table}[t]
    \centering
    \renewcommand{\arraystretch}{1.2}
    \addtolength{\tabcolsep}{5pt} 
    \begin{tabular}{l | l | l | l}
        \hline \hline
        \multirow{2}{*}{Model component}
         & \multicolumn{1}{c|}{on sphere} &  \multicolumn{2}{c}{at satellite radius}  
        \\ \cline{3-4}
        & \multicolumn{1}{c|}{same time} & \multicolumn{1}{c|}{same time} & \multicolumn{1}{c}{time dependent}
        \\
        \hline
        lithosphere & $-0.000033(\pm 24)$ & $-0.000087(\pm 26)$ & $-0.000087(\pm 26)$ \tablefootnote{The lithosphere component is time independent by model assumption.}
        \\
        external & $-0.0000088(\pm 66)$ & $\: \: \, \, 0.013\pm 0.000008$ & $\: \: \, \, 0.022\pm 0.0001$ 
        \\
        core dipole & $\: \: \, \, 0.032\pm 0.025$ & $\: \: \, \, 0.032\pm 0.025$ & $\: \: \, \, 0.030\pm 0.025$ 
        \\
        core higher moments & $-0.20\pm 0.01$ & $-0.56\pm 0.01$ & $-0.57\pm 0.01$
        \\
        \hline
        full model & $-0.17\pm 0.03$ & $-0.51\pm 0.03$ & $-0.52\pm 0.03$ 
        \\
        \hline \hline
    \end{tabular}
    \caption{Values for the average flux $\langle\overline{\mathscr{B}}^\text{mod}\rangle$ and its $1\sigma$ statistical error in nT from various model components in the CHAOS-7 model~\cite{Finlay_2020} of Earth's magnetic field, evaluated along the satellites' trajectories and spanning the time period from Feb 1, 2014 to June 29, 2020. The second and third columns remove the model's time dependence and respectively assume either the measurements are taken on a sphere with fixed radius or at the satellite radii and rescaled by $(r/R_\text{ref})^3$ as in (\ref{eq:B-modified}). The final column also uses the satellite radii and includes the model's time dependence. Results are for $a=2$, $d=180$, and $Kp \leqslant 3$. The four components do not sum to the full model as they are weight-averaged by their own variance, as discussed in Sec.~\ref{sec:stat}.
    }
    \label{tab:model}
\end{table}

The third column uses the actual variation in the satellite altitude and rescales the field by $(r/R_\text{ref})^3$ as in Eq.~(\ref{eq:B-modified}), but still removes time dependence. This shows the error introduced by the $r^3$ rescaling. Note, the dipole results match whether taken on a sphere or rescaled because the two operations are mathematically equivalent. The dipole dominates the full model's error because its overall contribution to the field is larger, and its value is consistent with zero. On the other hand, the higher core moments have much larger $\langle\overline{\mathscr{B}}^\text{mod}\rangle$ when calculated at the satellite radius (compared with both the dipole and higher moment calculation on sphere) because higher moments are proportional to $r^{-n}$ with $n>3$. Thus, higher moments provide the dominant contribution to $\langle\overline{\mathscr{B}}^\text{mod}\rangle$. Unlike the on-sphere calculation, there is relatively little suppression of $\langle\overline{\mathscr{B}}^\text{mod}\rangle$ as $a \to 0$ because of the saturated contribution from the higher moments when evaluated at the satellite radius.
Meanwhile, the mean external component evaluated at satellite radius turns out to have a large significance compared with its statistical error, but it contributes only a small amount to the full model prediction.

Finally, the last column shows the inclusion of time-dependence to the model, which is most relevant to our calculation. For most components, time dependence is a subdominant effect in that there is minimal change between the third and fourth columns. The exception is the external component, which is expected to have significantly more time variation than the terrestrial components on timescales relevant to the data.

\section{Results}

Values for $a$ and $d$ are chosen for numerical stability in the calculation of both $\langle\overline{\mathscr{B}}^\text{dif}\rangle$ and its uncertainty and to obtain good agreement between all time bins and the average value. These tend to prefer larger $d$ and smaller $a$, although $a$ cannot be too small so that all angular patches are measured within each time bin. In general, as shown in the Supplemental Materials, any value of $a$ between $2$ and $0.3$ give similar results for large enough $d$ (with the requisite $d$ increasing with decreasing $a$). They are also robust against the choice of $Kp$ cut. 

\begin{figure}[t!]
    \centering
    \includegraphics[width=0.6\textwidth]{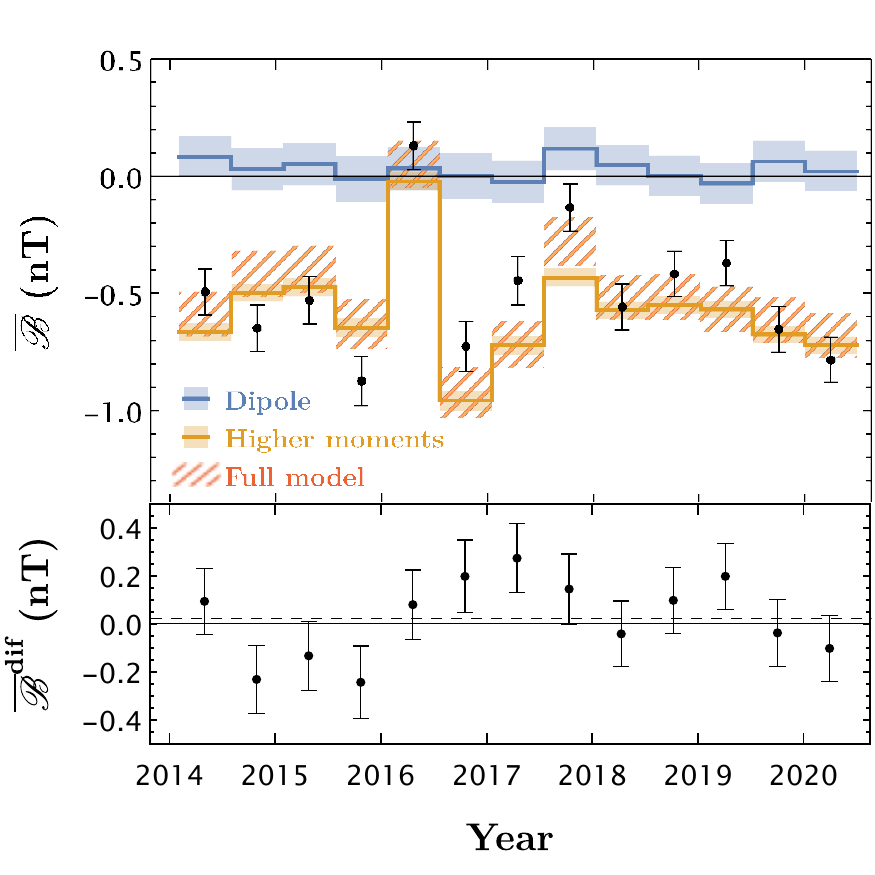}
    \caption{Average value and error of $\overline{\mathscr{B}}$ in 180-day bins using $2^\circ$ angular patch size covering the time period from 1 Feb 2014 to 29 June 2020. Data from {\it Swarm A} and {\it B} were incorporated, with the selection criteria $Kp\leqslant 3$. In the top panel, the blue and yellow lines indicate the contribution from the core's dipole and higher moments, respectively, to the model prediction, with the shaded bands giving their errors. The hatched regions show the total model predictions with errors. Data are shown by black points. In the lower panel, the differences between the data and model are shown, and the dashed line indicates the mean. All error bars are $1\sigma$ and include only statistical error.}
    \label{fig:result}
\end{figure}

Using $a = 2$ and $d = 180$, the average difference between the data and model is $\langle\overline{\mathscr{B}}^\text{dif}\rangle = 0.022 \pm 0.046~\text{nT}$, with a goodness of fit to the mean $\chi^2 = 16.2$ with $N_k-1=12$ degrees of freedom. The quoted uncertainty has been marginally enhanced by the Birge factor from the purely statistical uncertainty of 0.039 nT. Full sky coverage is obtained in each time bin.
A plot of the measurements in each time bin for both data and model is shown in Fig.~\ref{fig:result}, along with the contributions from the dipole and higher moments individually. The data and model follow each other very closely. The model's value is dominated by the core's higher moments, but its uncertainty is dominated by the core's dipole moment because of the dipole's much larger overall contribution to the magnetic field. If $a$ is decreased, as shown in the Supplementary Material, the statistical error bars for individual time bins shrink, but the overall error is dominated by the variation between time bins and levels off.

Because, no significant difference between the data and model is found, we perform a signal injection to set a bound on the magnitude of the monopole component field at 95\% confidence level: $-0.07~\text{nT} < B_\text{m}(r=R_\text{ref}) < 0.11~\text{nT}$, in agreement with the $2\sigma$ range of $\langle\overline{\mathscr{B}}^\text{dif}\rangle$.  This translates into an upper bound on the monopole component field at Earth's surface $|B_\text{m}(r=R_\oplus)|< 0.13~{\rm nT}$ and on the net magnetic charge $|Q_{\rm net}| < 1.6\times 10^{19} \equiv Q_\text{max}$. 

\section{Constraints on magnetic monopole objects}

One can apply the constraints on the total net magnetic charge $|Q_{\rm net}| < Q_{\rm max}$ into various fundamental theories that predict magnetic monopole objects with different charge-to-mass ratios, which we parametrize by $q \equiv Q/Q_\text{ext}$ with $Q_\text{ext} = e \, M / [\cos (\theta_W) \sqrt{\pi} M_\text{pl}] \approx 0.19 \, M/M_\text{pl}$. Here, $Q_\text{ext}$ is the charge of an extremal magnetic black hole with an electroweak symmetric corona of mass $M$ \cite{Maldacena:2020skw,Bai:2020spd}, with $M_\text{pl} = 1.22 \times 10^{19}~\GeV/c^2$ the Planck mass, $\theta_W$ the weak mixing angle, and $e$ the electric coupling constant. A magnetic black hole has $q \leqslant 1$ (saturated to equality in the extremal limit) and $M > M_\text{pl}$. Because magnetic black holes can efficiently Hawking radiate into electrons if their temperature is sufficiently large, they satisfy $q \sim 1$ whenever $M \lesssim 10^{17}~\text{g}$ \cite{Maldacena:2020skw,Bai:2020spd,Diamond:2021scl}, but can take on any $q \leqslant 1$ at larger masses. Conversely, a monopole particle has $q>1$ according to the weak gravity conjecture \cite{ArkaniHamed:2006dz}. While a GUT monopole with $Q=2$ and mass $M \simeq 10^{17}~\GeV/c^2$ has $q \simeq 1300$, a gravitating composite monopole  object with a large magnetic charge could have much larger mass with both small and large $q$~\cite{Hartmann:2000gx}.
Therefore, we treat $q$ and $M$ as free model parameters to set limits. 

\begin{figure}[t]
    \centering
    \includegraphics{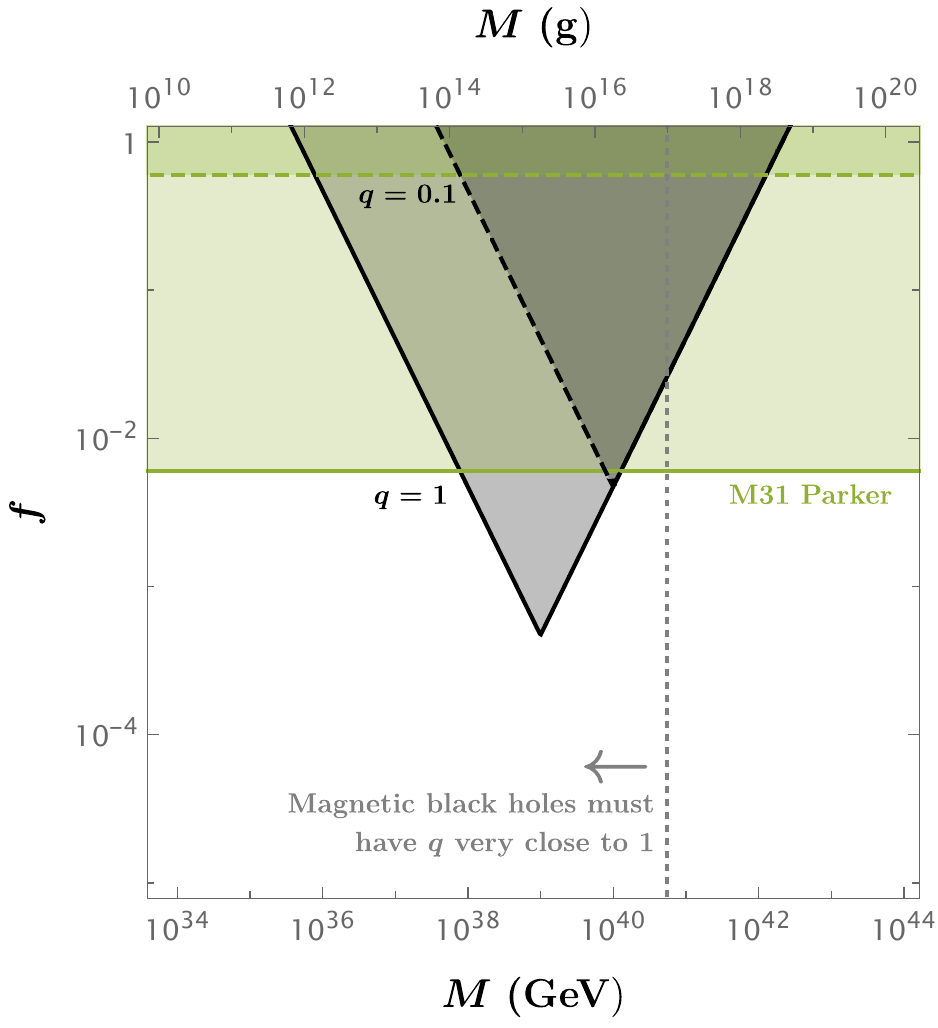}
    \caption{Bounds on the local energy density of magnetic monopoles as a fraction of the local dark matter density. Black shaded is the bound from the Earth's magnetic monopole moment, bounding Earth's net magnetic charge by $Q_\text{max}=1.6\times 10^{19}$. Green is the M31 Parker bound \cite{Bai:2020spd}. Solid is $q=1$, while dashed is $q=0.1$. To the left of the gray dashed line, magnetic black holes must be close to the stable extremal state ($q \sim 1$), though other magnetically charged objects with different $q$ could exist.}
    \label{fig:abundance}
\end{figure}

The capture rate of magnetic monopoles is estimated to be $C_{\rm cap} \approx \pi R^2_\oplus\,4\pi\,F$. Here, $F\approx (1.7\times 10^{-33}\,\mbox{cm}^{-2}\mbox{sr}^{-1}\mbox{s}^{-1}) \, f \, (10^{15}~\text{g}/M)$ is the magnetic monopole flux with $v\approx 10^{-3}\,c$ the averaged speed for a heavy monopole bounded in our galaxy and $f = \rho / (0.4 \,\mbox{GeV}\,\mbox{cm}^{-3})$ the local monopole energy density $\rho$ as a fraction of the local dark matter density~\cite{Bai:2020spd}. 
\footnote{All magnetic black holes incident on Earth with the large charges and masses considered here will be captured~\cite{Bai:2020spd}.}
Using the Earth's lifetime $\tau_\oplus \approx 1.4\times 10^{17}\,\mbox{s}$, the number of captured monopoles is $N=C_{\rm cap}\,\tau_\oplus\approx 3800 \, f \, (10^{15}~\text{g}/M)$. When $N \geq 1$, the net charge of captured monopoles is $Q_\text{net} \simeq \sqrt{N} Q$. Thus, the constraint $Q_\text{net} < Q_\text{max}$ can be expressed as a limit on the local density of monopoles: $f \lesssim 8.8 \times 10^{-4} \, q^{-2} \, [Q_\text{max}/(1.6 \times 10^{19})]^2 \, [10^{15}~\text{g}/M]$, valid in the regime $f \gtrsim 2.6 \times 10^{-4} \, [M/(10^{15}~\text{g})]$. This is depicted in Fig.~\ref{fig:abundance}. Also shown is the Parker bound \cite{Parker:1970xv,Turner:1982ag} derived from M31 in \cite{Bai:2020spd}, $f \lesssim 6 \times 10^{-3} \, q^{-2}$, which disappears when $q < 0.08$. The limit presented here is complimentary to other limits---for example from gas heating and white dwarf destruction \cite{Diamond:2021scl}---in that it is a direct measurement as opposed to an inference from difficult-to-model astrophysical systems.

\section{Discussion}

The Earth's magnetic field provides an interesting way to search for new physics. Using satellite-based measurements, we have shown it is possible to estimate a bound on the magnetic flux with a modified version of Gauss's law. Even without any knowledge of the higher multiple moments, the monopole moment can be constrained at the level of $\mathcal{O}(\text{nT})$ (the value for $\mathscr{\overline{B}}$ when $r^3$ rescaling as in (\ref{eq:B-modified}) is employed), and comparing to a model of the higher moments allows for bounds an order of magnitude stronger. This suggests that future model fits should include the possibility of a monopole term. Combining a global fitting technique with data from more observatories would provide the best achievable bound with present technology. We encourage future work in this direction to build on the work presented here. Mars, which also has satellite-based measurements of its magnetic field \cite{2015SSRv..195....3J}, could be another interesting target for future work.

\subsubsection*{Acknowledgements}
We thank Joshua Berger and Mrunal Korwar for useful discussion. The work of YB is supported by the U.S. Department of Energy under the contract DE-SC-0017647. The work of SL is supported in part by Israel Science Foundation under Grant No. 1302/19. The work of NO is supported by the Arthur B. McDonald Canadian Astroparticle Physics Research Institute.

\appendix
\section{Supplemental Materials}

\begin{figure}[ht!]
    \centering
    \includegraphics[width=0.6\textwidth]{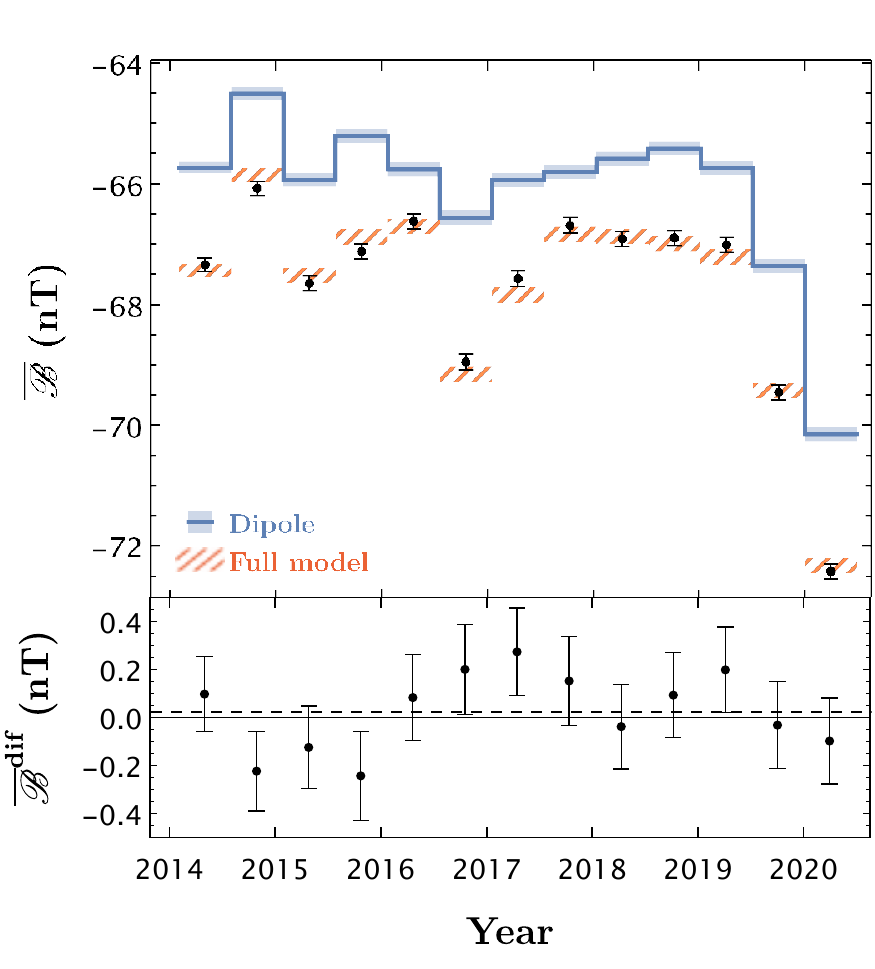}
    \caption{
    Same as Fig.~\ref{fig:result}, but $\overline{\mathscr{B}}$ is calculated without the $r^3$ rescaling in Eq.~(\ref{eq:B-modified}). All components other than the dipole have $|\overline{\mathscr{B}}|<3~\text{nT}$, far off the range of the plot but a bit larger compared to $|\overline{\mathscr{B}}|<1~\text{nT}$ when $r^3$ rescaling is employed. The central value for $\langle\overline{\mathscr{B}}^\text{dif}\rangle$ is close to zero as in Fig.~\ref{fig:result}, while the error bar size is larger.}
    \label{fig:no_rescale}
\end{figure}

In this Supplemental Material, we first address the effectiveness of the $r^3$-rescaling discussed in the main text. In Fig.~\ref{fig:no_rescale}, we plot the calculated time sequence of $\overline{\mathscr{B}}$ without $r^3$ rescaling using 180-day time bin, $2^\circ$ angular patch size, and $Kp\leqslant 3$. 
Shown in the upper panel are the $\overline{\mathscr{B}}$ from both model and data.
Both results have $\langle\overline{\mathscr{B}}\rangle \sim -70$~nT and significantly deviate from zero.
This value is dominated by the dipole component of the model, 
confirming that integrating $\bm{B}\cdot \mathbf{\hat{r}}$ does not provide a good estimate for the flux when the surface of integration is non-spherical. 
However, comparing the lower panels of Fig.~\ref{fig:result} and Fig.~\ref{fig:no_rescale}, it is clear that $\overline{\mathscr{B}}^{\rm dif}$ has a similar behavior in both, but the error bars are smaller when $r^3$ rescaling is used. Essentially, the $r^3$ rescaling has reduced the ``background'' contributed by the dipole, allowing us to set a stronger limit and thus justifying our use of the $r^3$ rescaling. The resulting average difference without $r^3$ rescaling is $\langle\overline{\mathscr{B}}^\text{dif}\rangle = 0.023 \pm 0.049~\text{nT}$, again consistent with the result in the main text but setting a less stringent bound on the monopole flux. As $a \to 0$, the statistical errors become less important, and the variation between time bins sets the error to be about $\sigma^\text{sem}$, comparable to that of the $r^3$-rescaled calculation. Still, the $r^3$ rescaling allows for easier comparison between the model and data.

\begin{figure}[ht!]
    \centering
    \includegraphics[width=\linewidth]{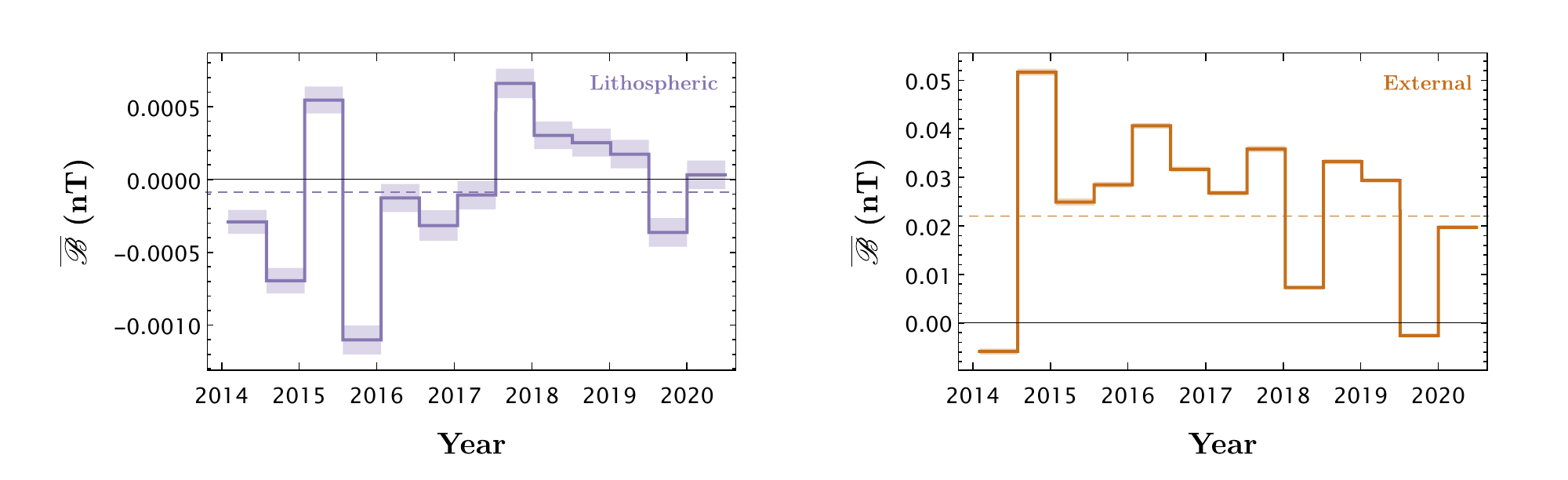}
    \caption{Average value and error of $\overline{\mathscr{B}}$ from the lithosphere components and external magnetic field in 180-day bins using $a=2$ covering the time period from 1 Feb 2014 to 20 Jun 2020, exactly as in Fig.~\ref{fig:result}. Dashed lines show their respective averages. Data are calculated with the CHAOS-7 model.}
    \label{fig:crust_vs_ext}
\end{figure}

Next, we show for completeness the time sequence of the lithosphere and the external components of the model for $d=180$ and $a=2$ (including $r^3$ rescaling) in Fig.~\ref{fig:crust_vs_ext}, complementary to Fig.~\ref{fig:result}.
The magnitude of the these components are smaller compared with the two core components and therefore do not play a significant role in the analysis, although they are always included.

In addition, in Table~\ref{tab:results} we show $\langle\overline{\mathscr{B}}^{\rm dif}\rangle$ for several combinations of $d$, $a$, and $Kp$ cut other than the benchmark combination used in the main text.

When the amount of data per angular patch is large enough, the results are insensitive to the choice of $a$ in the presented range when $d$ and the $Kp$ cut are fixed, indicating stability in the numerical integration. The errors (including Birge ratio factors) are also insensitive to $a$ as explained in Section~\ref{sec:stat}. 
When $d$ varies (with other variables fixed), the result fluctuates slightly because different amounts of data are included (2362 days of data are included in total, which is not always divisible by $d$, and the remainder are not used). 
Also, when $d$ becomes too small, not all time bins meet the 100\% sky coverage criteria, resulting in further fluctuations and larger error estimates. 
For results quoted in Table~\ref{tab:results}, only the $a=1$, $d=90$, and $Kp \leqslant 2$ cut is affected by this sky coverage criteria, and only one time bin is excluded on these grounds.
We also find the results to be largely insensitive to variation of the $Kp$ cut. 
On the other hand, we do not present results for $a<1$ because $\langle\overline{\mathscr{B}}^{\rm dif}\rangle$ and its error becomes numerically unstable, mainly due to the lack of data in each angular patch, though they may become stable if $d$ is large enough. 
For all the reasonable combinations of $d$, $a$, and $Kp$ we checked, $\langle\overline{\mathscr{B}}^{\rm dif}\rangle$ is consistent with 0 within $2\sigma$ errors.

\begin{table}[hb!]
  \renewcommand{\arraystretch}{1.2}
    \addtolength{\tabcolsep}{5pt} 
    \centering
    \begin{tabular}{c | c | c | c | c}
        \hline \hline
        $Kp \leqslant$ & $d$ (days) & $a= 2$ & $a= 1.5$ & $a= 1$ \\
        \hline
        \multirow{4}{*}{2} & 90 & $0.017\pm 0.052$ & $0.017\pm 0.052$ & $0.015\pm 0.060$\\
            & 120 & $0.039\pm 0.079$ & $0.042\pm 0.079$ & $0.039\pm 0.078$ \\
            & 180 & $0.020\pm 0.048$ & $0.020\pm 0.048$ & $0.019\pm 0.049$ \\
            & 330 & $0.027\pm 0.046$ & $0.028\pm 0.046$ & $0.028\pm 0.046$ \\
        \hline
        \multirow{4}{*}{3} & 90 & $0.020\pm 0.051$ & $0.020\pm 0.051$ & $0.019\pm 0.052$\\
            & 120 & $0.039\pm 0.079$ & $0.042\pm 0.079$ & $0.039\pm 0.078$ \\
            & 180 & $0.022\pm 0.046$ & $0.023\pm 0.046$ & $0.023\pm 0.047$ \\
            & 330 & $0.028\pm 0.048$ & $0.029\pm 0.048$ & $0.028\pm 0.048$ \\
        \hline \hline
    \end{tabular}
    \caption{Weighted mean of the difference between data and model and its statistical error for several different $a$, $d$ and $Kp$, calculated with method described in Subsection \ref{sec:stat} after including the Birge factor for the error. Though not shown in the table, for $d\geqslant 180$ days the statistical error of the mean is largely comparable.}
    \label{tab:results}
\end{table}

\begin{figure}[ht!]
    \centering
    \includegraphics[width=0.5\linewidth]{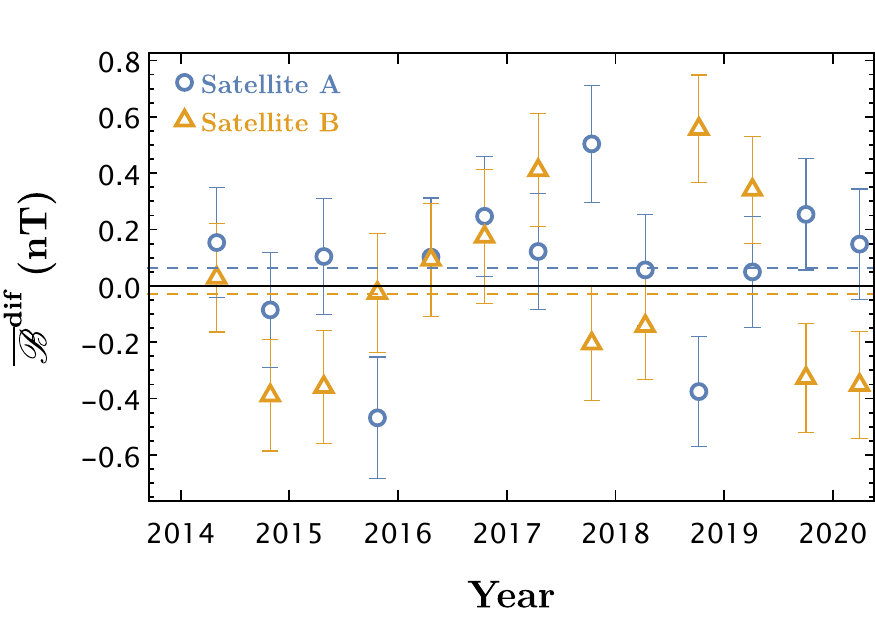}
    \caption{$\langle\overline{\mathscr{B}}^{\rm dif}\rangle$ and its statistical error calculated for {\it Swarm} satellite {\it A} and {\it B} separately, with $d=180$ using $a=2$ covering the time period from 1 Feb 2014 to 20 Jun 2020. The reference radius $R_{\rm ref}=$ 6820 and 6880 km for satellite {\it A} and {\it B} respectively. Dashed lines show their respective averages.
    }
    \label{fig:singleSatellite}
\end{figure}

Additionally, the data from both satellites {\it Swarm A} and {\it B} are largely consistent with each other. Using $a = 2$ and $d=180$, the calculated $\langle\overline{\mathscr{B}}^{\rm dif}\rangle$ in each time bin for each individual satellite are plotted in Fig.~\ref{fig:singleSatellite}. There, rather than using the average satellite radius for both satellites, we use each satellites' average radius: $R_\text{ref}=6820~\text{km}$ for {\it A} and $R_\text{ref}=6880~\text{km}$ for satellite {\it B}. The resulting averages are $\langle\overline{\mathscr{B}}^\text{dif}\rangle = 0.063 \pm 0.070~\text{nT}$ for {\it A} only and $\langle\overline{\mathscr{B}}^\text{dif}\rangle = -0.029 \pm 0.091~\text{nT}$ for {\it B} only. While the results for each satellite are in slight statistical tension with each other due to their different orbital paths, both are consistent with a null signal. The tension is reduced when the Birge factor is applied to each data set.

\setlength{\bibsep}{6pt}
\bibliographystyle{JHEP}
\bibliography{earthmonopole.bib}
\end{document}